\documentclass[sigconf]{acmart}
\acmConference[ESEC/FSE 2019]{The 27th ACM Joint European Software Engineering Conference and Symposium on the Foundations of Software Engineering}{26--30 August, 2019}{Tallinn, Estonia}
\acmYear{2019}

\usepackage{booktabs} 
\usepackage{amssymb,amsfonts,amsmath,subfigure,url}
\usepackage{graphicx}
\usepackage{color}
\usepackage{multirow}
\usepackage{balance}
\usepackage{amssymb,amsfonts,amsmath}
\usepackage{graphicx,epsfig,boxedminipage,url,xspace,array}
\usepackage{pstricks,pst-node,pst-tree}

\usepackage[latin1]{inputenc}
  {\noindent\textbf{Proof:}\newline\begin{sffamily}\small}%
  {\end{sffamily}\hfill {\framebox(5, 5)[!t]{}}}
\providecommand{\LyX}{L\kern-.1667em\lower.25em\hbox{Y}\kern-.125emX\@}

\makeatletter
\newcommand{\@jux}[1]{#1 \mskip-5.5mu #1}
\newcommand{\@jux@}[1]{#1 \mskip-2.6mu #1}

\newcommand{\mvB}[1]{\@jux@{[} #1 \@jux@{]}^B}
\@ifundefined{bigsqcap}
  {}
  {\relax}
\makeatother

\newcolumntype{H}{>{$(}r<{)$}}

\usepackage{makecell}
\usepackage{algorithm}
\usepackage{algorithmicx}
\usepackage[noend]{algpseudocode}
\usepackage{multirow}
\usepackage{placeins}

\usepackage{framed}
\definecolor{light-gray}{gray}{0.9}
\usepackage[framemethod=TikZ]{mdframed}
\mdfsetup{skipabove=5pt,skipbelow=3pt}
\usepackage{lipsum}
\mdfdefinestyle{MyFrame}{%
    linecolor=black,
    outerlinewidth=0.15pt,
    roundcorner=3pt,
    innertopmargin=2pt,
    innerbottommargin=2pt,
    innerrightmargin=4pt,
    innerleftmargin=4pt,
    backgroundcolor=light-gray}

\newboolean{showcomments}
\setboolean{showcomments}{true} 
\ifthenelse{\boolean{showcomments}}
{\newcommand{\nb}[2]{
  \fcolorbox{black}{yellow}{\bfseries\sffamily\scriptsize#1}
  {\sf\small$\blacktriangleright$\textit{#2}$\blacktriangleleft$}
 }
 
}
{\newcommand{\nb}[2]{}
 
}

\usepackage{changes}
\definechangesauthor[name={Khouloud Gaaloul},color=orange]{KG}
\definechangesauthor[name={Claudio Menghi},color=orange]{CM}
\definechangesauthor[name={ShivaNejati},color=orange]{SN}

\usepackage[colorinlistoftodos,prependcaption,textsize=tiny]{todonotes}

\begin{document}
\title[Evaluating Model Testing and Model Checking]{Evaluating Model Testing and Model Checking for Finding Requirements Violations in Simulink Models}

\author{Shiva Nejati}
\affiliation{%
  \institution{University of Luxembourg}
  \state{Luxembourg}
}
\email{shiva.nejati@uni.lu}

\author{Khouloud Gaaloul}
\affiliation{%
  \institution{University of Luxembourg}
  \state{Luxembourg}
}
\email{khouloud.gaaloul@uni.lu}

\author{Claudio Menghi}
\affiliation{%
  \institution{University of Luxembourg}
  \state{Luxembourg}
}
\email{claudio.menghi@uni.lu}

\author{Lionel Briand}
\affiliation{%
  \institution{University of Luxembourg}
  \state{Luxembourg}
}
\email{lionel.briand@uni.lu}

\author{Stephen Foster}
\affiliation{%
  \institution{QRA, Corp}
  \state{Canada} 
}
\email{stephen.foster@qracorp.com }

\author{David Wolfe}
\affiliation{%
  \institution{QRA, Corp}
  \state{Canada} 
}
\email{david.wolfe@qracorp.com}

\renewcommand{\shortauthors}{Nejati, Gaaloul, Menghi,  Briand, Foster and Wolfe}

\begin{abstract}
Matlab/Simulink is a development and simulation language that  is widely used by the Cyber-Physical System (CPS) industry to model dynamical systems. There are two mainstream approaches to verify CPS Simulink models: \emph{model testing} that attempts to identify failures in models by executing them for a number of sampled test inputs, and \emph{model checking} that  attempts to exhaustively  check the correctness of models against some given formal properties. In this paper, we present an industrial Simulink model  benchmark, provide a categorization of different model types in the benchmark, describe the recurring logical patterns in the model requirements,  and discuss the results of applying model checking and model testing approaches to identify requirements violations in the benchmarked models. Based on the results, we discuss the strengths and weaknesses of model testing and model checking. 
Our results further suggest that model checking and model testing  are complementary and by combining them, we can significantly enhance the capabilities of each of these approaches individually. We conclude by providing  guidelines as to how the two approaches can be best applied together.
\end{abstract}

\maketitle


\section{Introduction}
\label{sec:intro}

The development of Cyber Physical Systems (CPS) relies on early function modeling of the system and its environment. These models typically capture dynamical systems. For example, they may be mathematical models capturing movements of a physical object or they may specify a software controller that interacts with a physical object or a physical process to respectively control the movements of the object or the progression of the process over time. A key and common feature of these models is that they typically consist of time-varying and real-valued variables and functions.  

Matlab/Simulink is a development and simulation language that  is widely used by the CPS industry to capture CPS dynamical systems. Specifically, Simulink is used by more than 60\% of engineers for simulation of CPS~\cite{zheng2017perceptions,baresi2017test}, and is the prevalent modeling language in the automotive domain~\cite{Matinnejad:2016:ATS:2884781.2884797,zander:12}. Simulink appeals to engineers since it is particularly suitable for specifying mathematical models and dynamic systems, and further, it is executable and allows engineers to test their models as early as possible.

To avoid ripple effects from defects and to ensure that failures are identified as early as possible, it is paramount for the CPS industry to  ensure that CPS Simulink models satisfy their functional safety requirements.   Different approaches to verification and testing of Simulink models have been proposed in the literature~\cite{hamon:08,barnat:12,matinnejad2018test,Abbas:13}. The majority of them fall under one of the following  two main categories: (1)~\emph{Model checking} techniques that attempt to  exhaustively verify the correctness of models against some given formal properties~\cite{Clarke1999}. (2)~\emph{Model testing} techniques that attempt to identify failures in models by executing them for some test inputs sampled by a guided randomized algorithm~\cite{BriandNSB25,Abbas:13,matinnejad2018test}. Model checking approaches often translate Simulink models as well as the given properties into the input language of some existing model checkers or Satisfiability Modulo Theories (SMT) solvers. The main disadvantage of model checking when applied to Simulink models is that these models often capture continuous dynamic and hybrid systems~\cite{alur:15}.  It is well-known that model checking such  systems is in general undecidable~\cite{henzinger1998s,alur1995algorithmic,6064535}. The translation of continuous systems into discrete logic often has to be handled on a case-by-case basis and involves loss of precision which may or may not be acceptable depending on the application domain.  Further, industrial Simulink models often contain features and constructs that cannot easily be translated into low-level logic-based languages. Such features include  third-party code (often encapsulated in Matlab S-Functions) and non-algebraic arithmetics (e.g., trigonometric, exponential, and logarithmic functions).  Nevertheless, model checking, when applicable,  can both detect faults and demonstrate lack thereof. Model testing, on the other hand, due to its black-box nature, does not suffer from such applicability and scalability issues, but it can only show the presence of failures and not their absence. The effectiveness of model testing techniques highly relies on their underlying heuristics and search guidance strategies. Since there is no theoretically proven way to assess and compare different search heuristics, model testing techniques can only be evaluated  empirically.

\begin{figure}[t]
	\centering
	\includegraphics[width=1\columnwidth]{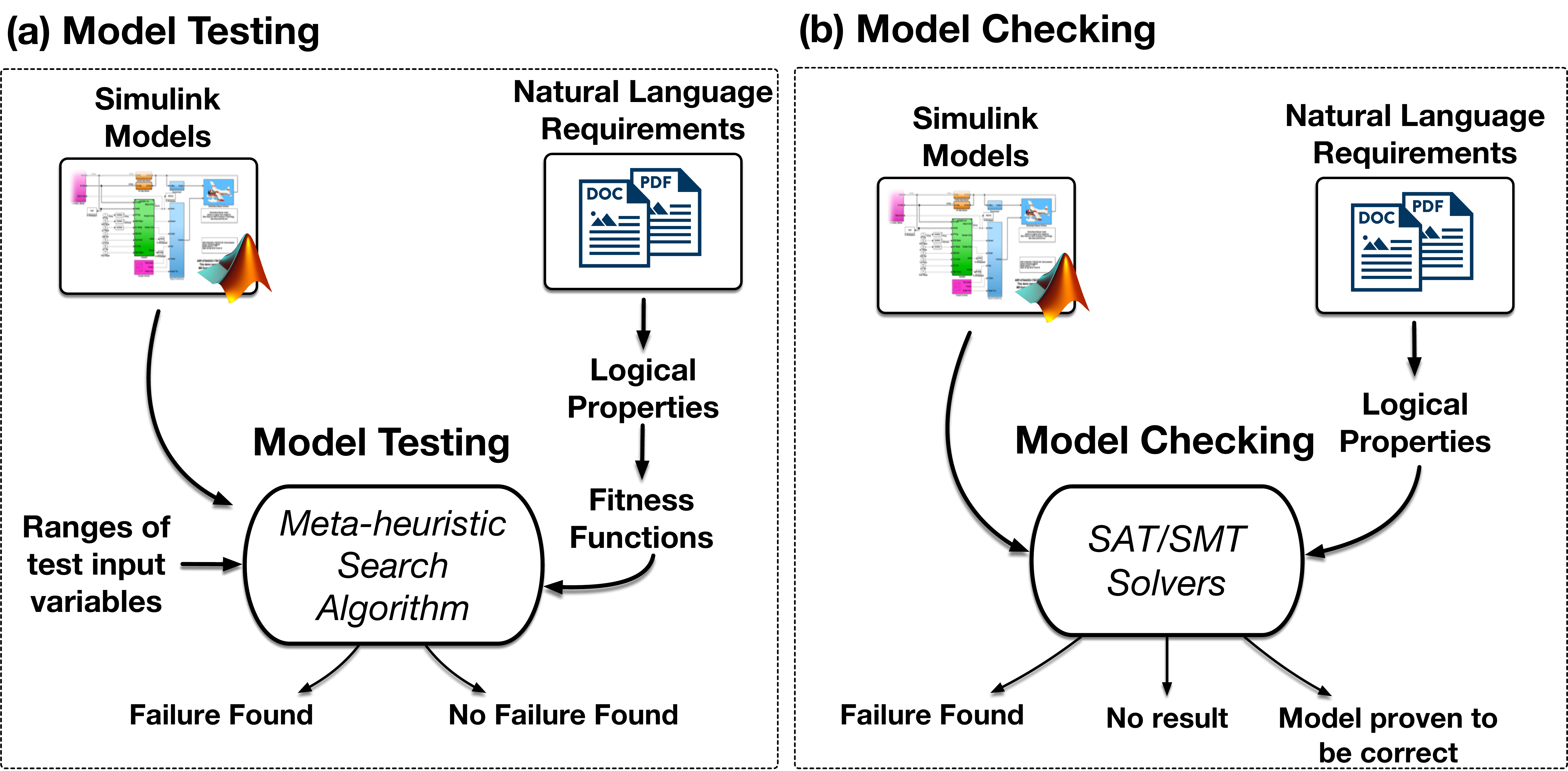}
	\caption{Simulink Model Verification: (a)~Model testing and (b)~Model checking.}
	\label{fig:mc-mt}
\end{figure}

In addition to model checking and model testing, \emph{statistical model checking} has also been previously applied to Simulink models~\cite{younes:06,legay:10}. This technique aims to provide probabilistic guarantees, subject to assumptions about the distribution of system inputs,  indicating that a model satisfies its given formal properties~\cite{younes:06,zuliani:13}.  Specifically, statistical model checking uses uniformly sampled execution traces generated by the model under test together with statistical inference methods to determine whether the sampled traces provide a statistical evidence for the satisfaction of the properties of interest~\cite{zuliani:13,clarke:11}. Similar to model testing, statistical model checking has a black-box nature. However, in contrast to both model testing and model checking that can be used in early verification to find failures,  statistical model checking is targeted towards  software validation activities and produces probabilistic estimates about the correctness of models.

Despite the large volume of academic research on software testing and verification, there are relatively few commercial and industry-strength tools for testing and verification of Matlab/Simulink models. In particular, we identify three major commercial tools for testing Matlab/Simulink models: Reactis~\cite{reactistest, cleaveland:08}, Simulink Design Verifier (SLDV)~\cite{sldv} and QVTrace~\cite{qvtrace}.  Among these tools, Reactis  combines a guided random search strategy with coverage-based SMT model checking. Specifically, Reactis first generates test inputs randomly. Then, the coverage goals that are not covered by the randomly generated inputs are attempted to be covered using SMT-based model checking.  SLDV and QVTrace, on the other hand,  are SMT-based model checkers designed for Simulink models. Specifically, formal properties together with the models are translated into logical constraints that can be fed into SMT solvers. The SMT solvers then attempt to verify that given formal properties hold on the models,  or otherwise, they generate  counter-examples demonstrating the presence of faults in the models.

In this paper, we report on an empirical study evaluating capabilities of model testing and model checking techniques in finding faults in Simulink models. Specifically, we use a benchmark consisting of Simulink models from the CPS industry in our empirical study to compare the two approaches. The benchmark is developed by a well-known and major aerospace and defense  company (hereafter referred to as company A [real name redacted due to NDA]). The benchmark includes eleven Simulink models that are representative of different types of CPS behavioral models in the aerospace and defense sector.  Each model is accompanied by a set of functional requirements described in natural language that must be satisfied by the model. Each model further includes some faults that violate some of the model requirements. The faults in the models are introduced by company A based on their past experiences of common faults in CPS behavioral design models. Without revealing the locations and types of faults in the models, company A uses this benchmark to assess the capabilities of different verification and testing tools in the market. Testing tool vendors are requested to identify as many requirements violations as possible  when provided with the benchmark. The benchmark is available online~\cite{AdditionalMaterial}.

The model testing technique in our study builds on our prior work in this area~\cite{matinnejad2018test} and is implemented as a typical search-based testing framework~\cite{McMinn2004}. An overview of this framework is shown in Figure~\ref{fig:mc-mt}(a). In this framework, meta-heuristic search algorithms~\cite{Luke2013}  are used to explore the test input space and to select the best test inputs, i.e., the test inputs that reveal or are close to revealing requirements violations. The search is guided by fitness functions that act as distance functions and estimate how far test inputs are from violating a certain requirement.  In this paper, we use a search algorithm based on Hill Climbing heuristic~\cite{Luke2013} that, in prior work~\cite{MatinnejadNBBP15},  has shown to be effective in testing Simulink models.  We define search fitness functions using existing translations of logical formulas into quantitative functions estimating degrees of satisfaction of the formulas~\cite{Abbas:13}.

Among the commercial model checking tools for Simulink models (i.e., QVTrace and SLDV), we use the QVTrace tool~\cite{qvtrace} in our comparison. QVTrace is a recent commercial tool that builds on the ideas from SMT-based model checking.  SLDV, similar to QVTrace, is a SMT-based model checker. However, we chose to compare with QVTrace since the MathWorks license prevents publication of empirical results comparing with SLDV or any other MathWorks products. In contrast to SLDV, QVTrace is a standalone product developed by QRA~\cite{QRAcorp},  a Canada-based company specializing in the development of enterprise tools for early-stage validation and verification of critical systems. We further note that QVTrace is more recent than SLDV, has a wider range of features and benefits from a well-designed and usable interface. In contrast to SLDV that can only be used with Prover~\cite{prover}, QVTrace can be combined with various well-known SMT solvers and theorem provers such as Z3~\cite{DeMoura:2008:ZES:1792734.1792766} and Mathematica~\cite{mathematica}.

Our paper presents the following main results: 

\begin{itemize}

\item We provide a categorization of CPS model types and a set of common logical patterns in CPS functional requirements.  Specifically, using our industrial benchmark Simulink models, we identify a categorization of the CPS models based on their functions.  We further formalize the textual requirements in a logic-based requirements language and identify some common patterns among the CPS requirements in the benchmark.

\item We present the results of applying our model testing and model checking techniques to the Simulink benchmark. We evaluate  the fault finding abilities of both techniques.  This is a first attempt in the context of CPS, to systematically compare model checking and model testing -- the two main alternatives for verifying Simulink models -- on an industrial benchmark. The complete replication package for our study is available online (see Section~\ref{subsec:data})~\cite{AdditionalMaterial}.

\item We provide some lessons learned by outlining  the strengths and weaknesses of model testing and model checking in identifying faults in Simulink models. As these two approaches provide complementary benefits, we believe that integrating them in a comprehensive verification framework can result in an effective testing framework.
We further propose some guidelines as to how the two approaches can be best applied together.
 Finally,  we describe some directions for future work in this area. 
\end{itemize}

\emph{Organization.} Section~\ref{sec:cps-mod-req} presents our Simulink model benchmark, our CPS model categorization and our CPS requirements patterns. Section~\ref{sec:mc} summarizes the working of QVTrace, i.e., the model checking tool used in our study. Section~\ref{sec:mt} describes our model testing approach. Section~\ref{sec:eval} presents our empirical results. Section~\ref{sec:ll} discusses some lessons learned. Section~\ref{sec:con} concludes the paper.

\section{Simulink Benchmark}
\label{sec:cps-mod-req}
In this section, we present the CPS Simulink models and the CPS requirements in our Simulink benchmark. Table~\ref{table:cpsmodels} shows a summary of the models in the benchmark. In Section~\ref{subsec:exam}, we present two example models from the benchmark in more detail. In Section~\ref{subsec:model}, we categorize the benchmark models based on their functions. In Section~\ref{subsec:cpsreq}, we describe the logic language to formalize our CPS requirements and  present a number of recurring logic-based patterns in the requirements formalizations. 

\begin{table*}[t]
\caption{Important characteristics of our benchmark Simulink models (from left to right): (1)~model name, (2)~model description, (3)~model type (see Section~\ref{subsec:model}), and (4)~number of atomic blocks in the model. } 
\label{table:cpsmodels}
\begin{center}
\scalebox{.7}{\begin{tabular}{|p{0.2\linewidth}|p{0.75\linewidth} | p{0.3\linewidth}|p{0.06\linewidth}| }
 \toprule             
 \bf Model Name  & \bf Model Description & \bf Model Type & \bf \#Atomic Blocks \\
 \toprule  
	Autopilot & Discussed in Section~\ref{subsec:exam}. & Feedback-loop, continuous controller, plant model, non-linear, non-algebraic, matrix operations  & 1549 \\ \hline     
	\midrule 
 	Neural Network & A two-input single-output  predictor neural network model with two hidden layers arranged in a feed-forward neural network architecture. & Open-loop, machine learning& 704 \\\hline
	 \midrule
	Tustin & A numeric model that computes integral over time. &  Open-loop, non-linear (saturation and switches) & 57  \\\hline
 	\midrule 	
	Regulators& A PID controller without the plant model. &  Open-loop, continuous controller, non-linear (saturation, switches)& 308\\\hline
 	  	\midrule
  		Nonlinear Guidance &  A non-linear guidance algorithm that guides an Unmanned Aerial Vehicles (UAV) to follow a moving target respecting a specific safety distance. & Open-loop, non-linear (polynomial, switches) &  373 \\\hline
	\midrule
	System Wide Integrity Monitor (SWIM)& A numerical algorithm that computes warning to an operator when the airspeed is approaching a boundary where an evasive fly up manoeuvre cannot be achieved. &   Open-loop, non-linear (sqrt, switches) & 164  \\\hline
 	\midrule 
	Effector Blender & A control allocation method, which enables the calculation of the optimal effector configuration for a vehicle. &  Open-loop, non-linear (polynomial), matrix operations, non-algebraic (exponential functions) & 95 \\\hline
 	\midrule 
  	Two Tanks  & Discussed in Section~\ref{subsec:exam}.& 	Feedback-loop, sate machine, non-linear (switches)   & 498  \\\hline
  	\midrule
  	Finite State Machine (FSM)  & A finite state machine executing in real-time. Its main function is to  put the control of aircraft in the autopilot mode if a hazardous situation is identified in the pilot cabin (e.g., the pilot not being in charge of guiding the airplane) & Open-loop, sate machine, non-linear (switches) & 303 \\\hline
	\midrule 
	Euler & An open-loop mathematical  model that generates three-dimensional rotation matrices along the z-y- and x-axes of an Inertial frame in a Euclidean space. & Open-loop, non-algebraic (trigonometry),  non-linear (polynomial), matrix operations & 834   \\\hline
	\midrule 
	Triplex  & A monitoring system that receives three different sensor readings from three redundant sensors used in a safety critical system. It determines, based on the values and differences of three values received at each time step,  which sensor readings are trusted and what values should be sent to the safety critical system. & Open-loop, state machine, non-linear (switches) & 481 \\
  	\bottomrule

\end{tabular}}
\end{center}
\end{table*}

\subsection{Example Models}
\label{subsec:exam}
We highlight two example Simulink models from the benchmark: \emph{Two-Tanks}~\cite{GrossFH16}  and \emph{Autopilot}. These two models represent two complete systems instead of components of a system.  The two-tanks system  contains two separate tanks holding liquid  and connected via a pipe. The flow of incoming liquid into the first tank is controlled using a pump. The flow of liquid from the first tank to the second is controlled using a valve, and the flow of outgoing liquid from the second tank is controlled using two different valves: one that lets  liquid out in normal situations, and the other that is opened only in emergency conditions to avoid  liquid overflow. The model of the two-tanks system includes one controller model for each tank that monitors the liquid height using three different sensors located at different heights in each tank. Depending on the liquid heights, each controller chooses to open or close valves to control the incoming/outgoing liquid flows. 
The two-tanks model further includes a complete model of the environment (i.e., the tanks and their sensors and actuators). 

The autopilot system is a full six degree of freedom simulation of a single-engined high-wing propeller-driven airplane with autopilot. A six degree of freedom simulation enables movement and rotation of a rigid body in three-dimensional space. The autopilot simulator model is able to control the plane body to change position as  forward/backward (surge), up/down (heave) and left/right (sway) in three perpendicular axes, combined with changes in orientation through rotation in three perpendicular axes, often termed yaw (normal axis), pitch (transverse axis) and roll (longitudinal axis). The autopilot model further captures a physical model of the airplane (i.e., a plant model) as well as environment aspects impacting airplane movements such as wind speed. 

Both two-tanks and autopilot models use closed-loop controllers. However, the two-tanks controllers are modelled as discrete state machines, while the autopilot model consists of six continuous PID controllers~\cite{nise:04}. Some requirements of both models are described in Table~\ref{tab:requirements}. 

\begin{table*}[ht]
\caption{Example requirements  for the TwoTanks and Autopilot models.}
\label{tab:requirements}
\scalebox{.8}{\begin{tabular}{p{2cm}l p{10cm} | p{8.5cm}}
\toprule
\textbf{Model} &\textbf{ID} & \textbf{Requirement} & \textbf{Signal Temporal Logic formula (STL) *} \\
\toprule
{\bf Two Tanks} & {\bf R1} &If the liquid height of the first tank is greater than or equal to the position of the top sensor of the first tank,  then the top sensor should return an active (TRUE) state to the system. & $G_{[0,T]} (\mathit{tank1height} \geq \mathit{tank1topSensor} \Rightarrow \mathit{tank1sensorHValue = 1})$\\
\midrule
{\bf Two Tanks} &{\bf R2} &When the tank 2 MID sensor is TRUE, the tank 2 HIGH sensor is FALSE, and the emergency valve was previously open, then the emergency valve and the production valve (outflow valves) shall be commanded to be OPEN.& $G_{[0,T]}( ( \mathit{tank2sensorMValue} = 1 ) \wedge  ( \mathit{tank2sensorHValue} = 0 )\wedge  ( \mathit{eValveStatePrev} = 1 )  \Rightarrow  ( \mathit{eValveState} = 1 )\wedge ( \mathit{pValveState} = 1 )) $\\
\midrule
{\bf Autopilot} &{\bf R1} & The controller output will reach and stabilize at the desired output value within \%1 precision and within $T$ seconds (steady-state requirement). &  $F_{[0,T]} G_{[0,T]}  | \mathit{out} - \mathit{desired}| \le 0.01$\\
\midrule
  {\bf Autopilot} &{\bf R2} & Once the difference between the output and the desired value reaches  less than \%1, this difference shall not  exceed 10\% overshoot.   & 
 $G_{[0,T]} (|\mathit{out} -\mathit{desired}| \leq 0.01 \Rightarrow G_{[0,T]} |\mathit{out} -\mathit{desired}| \leq 0.1)  $\\

\bottomrule
\end{tabular}} 
\flushleft{\hspace*{.7cm}\small * Described in Section~\ref{subsec:cpsreq}.  The variable $T$ indicates the simulation time.}
\end{table*}


\subsection{CPS Simulink Models Categorization}
\label{subsec:model}
In the CPS domain, engineers use the Simulink language to capture  \emph{dynamic systems}~\cite{alur:15}.  Dynamic systems are usually used to model controller components as well as external components and the environment aspects that are to be controlled. The latter components are typically referred to as \emph{plants}. Dynamic systems' behaviors vary over time,  and hence, their inputs and outputs are represented as signals (i.e., functions over time). We describe some common categories of dynamical system components that we have  identified based on our industrial benchmark as well as  Simulink models from  other industrial sources~\cite{Matinnejad:17}.  We divide models into two large categories of \emph{open-loop} and  \emph{feedback-loop} models: 

\begin{figure}[t]
	\centering
	\includegraphics[width=1\columnwidth]{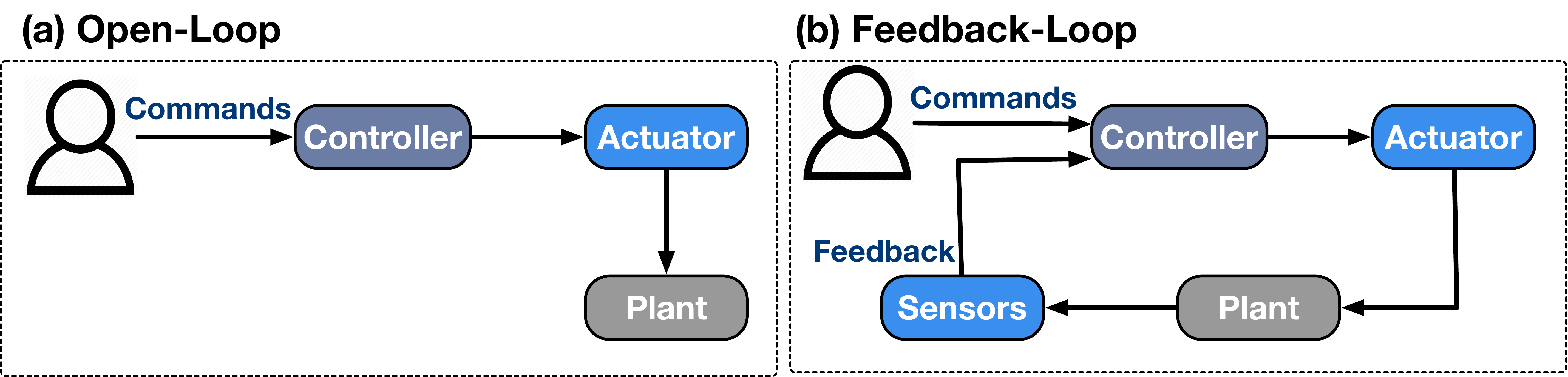}
	\caption{Generic structure of (a) open-loop and (b) feedback-loop CPS models}
	\label{fig:open}
\end{figure}

\vspace*{.2cm}
\emph{(1) Open-loop models} do not use measurements of the states or outputs of plants to make decisions~\cite{alur:15}. For example, an electronic cloth dryer  controller that relies on time to change its states is an open loop model. The user sets a timer for the controller, and the dryer will automatically stop  at the end of the specified time, even if  the clothes are still wet. The design of such controllers heavily relies on the assumption that the behavior of the plant is entirely predictable or determined.   Figure~\ref{fig:open}(a) represents the generic structure of open-loop models.

 \emph{(2)~Feedback-loop models} use measurements of the outputs or the states of plants to make decisions~\cite{alur:15}. 
This is the most common case in practical applications where engineers  need to design system controllers that act on some controlled inputs depending on the current state of the plants. For example, an electronic cloth dryer  controller that is able to stop when the clothes are sufficiently dried without requiring the user to set a timer, works by continuously monitoring the status of the clothes to choose when to stop the dryer. Such controllers are more flexible and are better able to handle unpredictable environment situations and  disturbances. Figure~\ref{fig:open}(b) represents the generic structure of closed-loop models. 
\vspace*{.15cm}

CPS models, whether being open-loop or feedback-loop, may consist of several components conforming to one or more of the following computation types: 

\textit{State machines.} State machines are used for modeling  discrete and high-level controllers. They can be used to monitor system behaviour or to control the system either in an open loop or closed loop model. Three models in our benchmark use state machines: (1)~Triplex is implemented using  a state machine to monitor three different sensor readings from three redundant sensors  and  identifies errors in the sensor readings; (2)~FSM uses an open-loop state machine controller to automatically put the control of aircraft in the autopilot mode if  a hazardeous situation is identified in the pilot cabin;   (3)~Two Tanks (discussed in Section~\ref{subsec:exam}) is  implemented as a composition of two state machine controllers for each tank arranged in a feedback-loop architecture together with physical models of the two tanks. Each state machine  controls pumps and valves of one tank. Since the pumps and valves can  only have two states (i.e., on and off), they can be controlled using state machines with a few states. In general, state machines are useful to model systems that have a clearly identifiable set of states and transitions that, when fired, change the state of the system.

\textit{Continuous behaviors.}  Continuous mathematical models are used to  describe both software controllers and physical plants. Continuous controllers,  also known as  \emph{proportional-integral-derivative} PID-controllers~\cite{nise:04}, are suitable when we need to control objects or processes whose states continuously vary over time. PID controllers are often used to control speed and movements of objects operating in varying environments with unpredictable disturbances. For example, the autopilot controller in Table~\ref{table:cpsmodels} contains six PID controllers. Plant models, which are  required for simulation of feedback-loop controllers, are typically described using continuous mathematical models. Continuous operations used in these two categories of models may have to be replaced with their discrete counterparts before the models can be translated into logic-based languages so that they can be analyzed by SMT-solvers.  Though continuous controllers also need to be discretized for code generation purposes, this is not the case of plant models, and therefore, discretization of plant models  is clearly an additional  overhead.

\textit{Non-linear and non-algebraic behaviors.}  CPS Simulink models are typically highly numeric. They often exhibit non-linear behavior or may contain non-algebraic functions, making their analysis complicated. In particular, the following operations make Simulink models non-linear: saturation blocks, switches, polynomial and square-root  functions, and the following operations are non-algebraic and are not typically supported by  SMT-solvers: trigonometry functions,  exponential functions and the logarithm. Finally, matrix operations are very commonly used in CPS Simulink models and are well-supported by Matlab. SMT-solvers, however, often do not directly support matrix operations, and hence, these operations have to be encoded and expanded  as non-matrix formulas. Therefore,  the size of the translations of Simulink models containing such operations into SMT-based languages become considerably larger than the size of the original models.

\textit{Machine learning models (ML).} Machine learning models are often used at the perception layer of dynamical systems (e.g., for image processing) or  are used to make predictions about certain behaviors. Verification of models inferred by machine learning techniques (e.g., Neural Networks) is an open area for research, as exhaustive verification techniques have only been applied to relatively simple and small neural network models~\cite{KatzBDJK17}. As shown in Table~\ref{table:cpsmodels}, we had one simple example machine learning component in our benchmark.

Table~\ref{table:cpsmodels} describes the types of the models in the benchmark by specifying whether they are open-loop or feedback-loop and also by indicating what component or feature types are used in each model. We note that most models are specified as open-loop since they are in fact sub-systems of a larger system that may have a feedback-loop architecture.  We note that the computation types  described above are not meant to be exhaustive.  Nevertheless, our categorisation provides more detailed information about the functional and behavioral variations in CPS models. Further, in Section~\ref{sec:eval}, we present the results of applying model checking and model testing approaches to our benchmark models, and the  categorisation  can further help determine how model checking and model testing approaches can deal with different computation types.

\subsection{CPS Requirements and Patterns}
\label{subsec:cpsreq}
As shown in Figures~\ref{fig:mc-mt}(a) and (b), both model checking and model testing require mathematical representations of requirements.  Specifically, model checking expects requirements to be described in temporal or propositional logic, and model testing expects them to be captured as quantitative fitness functions.  Requirements are properties the system must satisfy and usually constrain inputs and outputs behaviors. For CPS, model inputs and outputs are signals (i.e., functions over time). Therefore, the language used to formalize CPS requirements has to be defined over signal variables. Let $[a,b]$ s.t. $b \geq a \geq 0$ be an interval of real numbers. We denote signals by $s$ and define them as  $s:[a,b] \rightarrow \mathcal{R}$ where $\mathcal{R}$ is the signal range that can be boolean, enum or an interval of real numbers. We denote signals with a boolean or enum range by $s^\mathbb{B}$ and those with real intervals or  numbers by $s^\mathbb{R}$. In this paper, to formally specify model requirements, we use  Signal Temporal Logic (STL)~\cite{maler2004monitoring}, an extension of the well-known Linear Temporal Logic (LTL)~\cite{pnueli1977temporal} with \emph{real-time} temporal operators and \emph{real-valued} constraints. The syntax of STL is given below. 

\begin{center}
$\begin{array}{l}
\varphi ::=  \bot \mid \top \mid  s^{\mathbb{B}} \mid \mu \; \mathbf{rel\mbox{-}op} \; 0 \mid  \varphi_1 \vee \varphi_2 \mid  \varphi_1 \wedge \varphi_2 \mid \varphi_1 U_{[a,b]} \varphi_2 \\
\mu ::=   \mathbf{n} \mid s^{\mathbb{R}} \mid  \mu_1 \; \mathbf{math\mbox{-}op} \; \mu_2 \mid f(\mu) \mid (\mu)  \\
\end{array}$
\end{center}

where $s^{\mathbb{B}}$ is a boolean-valued signal, $s^{\mathbb{R}}$ is a real-valued signal,  $\mathbf{rel\mbox{-}op}$ is a relational operator (i.e., $\geq$, $>$, $<$, $\leq$, $=$, $\neq$), $\mathbf{math\mbox{-}op}$ is a numeric operator  (i.e., $+, -, \times, /$), $\mathbf{n}$ is a positive real number ($\mathbb{R}_0^+$) and $U_{[a,b]}$ is a real-time temporal operator. In the above grammar,  $f$ indicates a mathematical function applied to $\mu$ such as logarithm or trigonometry functions.

The semantic of STL is described in the literature~\cite{maler2004monitoring}. Briefly,  $\varphi$ formulas, except for $(\mu \; \mathbf{rel\mbox{-}op} \; 0)$, are temporal logic formulas where $U$ is the temporal until operator. In STL, the temporal until operator is augmented with an interval $[a,b]$ of real numbers  indicating that the until operator is applied to the signal segment between time instants $a$ and $b$.   Finally, the temporal eventually operator $F$ can be written based on the until operator as follows: $F_{[a,b]} \varphi = \top U_{[a,b]}  \varphi$, and the globally operator $G$ can be written as $G_{[a,b]}  \varphi = \neg F_{[a,b]}  \neg \varphi$. Note that when we write a temporal operator without specifying a time interval, we assume that the operator applies to the time interval of its underlying signals. For example, suppose we have signals defined over time interval $[0,T]$, we then write $G \varphi$ as a shorthand for $G_{[0,T]} \varphi$. 




\begin{table}[t]
\caption{Translation of Signal Temporal Logic~\cite{maler2004monitoring} into quantitative fitness functions to be used in the model testing approach  in Figure~\ref{fig:mc-mt}(a)}
\label{tbl:fitness}
\begin{center}
\scalebox{0.75}{\begin{tabular}{|p{0.6\linewidth}|p{0.6\linewidth}|}
\hline
 \multicolumn{2}{|c|}{\bf{Translation to robustness metric~\cite{fainekos2009robustness,Menghi:19}}}\\
\hline
\hline
 $R_{(S, t)}(\top) = \epsilon$ & $R_{(S, t)}(\bot) = -\epsilon$\\
\hline
$R_{(S, t)}(s^\mathbb{B})  =  \begin{cases} \epsilon&\mbox{if}\  s^\mathbb{B}  \\
	                         -\epsilon & \mbox{if}\ \neg s^\mathbb{B}\\
	                           	                          	                      
	     \end{cases}$ & $R_{(S, t)}(\mu = 0 )  =   - |\mu(S_t)|$  
	    \\
	    \hline
	  $R_{(S, t)}(\mu \neq 0 )  =   \begin{cases} |\mu(S_t)| &\mbox{if}\ \mu(S_t) \neq 0\\
	                   -\epsilon  & \mbox{else}\\
                               \end{cases}$  & 
$R_{(S, t)}(\mu \geq 0 ) = \mu(S_t) $ \\
\hline                   
$R_{(S, t)}(\mu > 0 )   =  \begin{cases} \mu(S_t) &\mbox{if}\ \mu(S_t) \neq 0\\
	                    - \epsilon   & \mbox{else}\\
                               \end{cases}$ & 
$R_{(S, t)}(\mu \leq 0 ) = - \mu(S_t) $ \\
\hline   
$R_{(S, t)}(\mu < 0 )   =  \begin{cases} -\mu(S_t) &\mbox{if}\ \mu(S_t) \neq 0\\
	                    - \epsilon & \mbox{else}\\
                               \end{cases}$ &
 \\
\hline
 \multicolumn{2}{|l|}{ $R_{(S, t)}(\varphi_1 \vee \varphi_2)  =  \mbox{max}(R_{(S, t)}(\varphi_1), R_{(S, t)}(\varphi_2))$} \\
  \hline
  \multicolumn{2}{|l|}{$R_{(S, t)}(\varphi_1 \wedge \varphi_2)  =  \mbox{min} (R_{(S, t)}(\varphi_1), R_{(S, t)}(\varphi_2))$}  \\
 \hline
 \multicolumn{2}{|l|}{  $R_{(S, t)}(G_{[a,b]}\varphi) = \mbox{min} \big\{ R_{(S, t')}(\varphi) \big\}_{t' \in [t+a, t+b]}$}\\
 \hline
  \multicolumn{2}{|l|}{$R_{(S, t)}(F_{[a,b]} \varphi) = \mbox{max}\big\{ R_{(S, t')}(\varphi)\big\}_{t' \in [t+a,t+b]} $} \\
\hline
 \multicolumn{2}{|l|}{$R_{(S, t)}(\varphi_1 U_{[a,b]} \varphi_2)  =  \mbox{max} \big\{\mbox{min}\big\{ R_{(S, t')}(\varphi_2), \mbox{min} \{R_{(S, t'')}(\varphi_1)\}_{t'' \in [t, t']}\big\} \big\}_{t' \in [t+a, t+b]}$} \\
\hline
\end{tabular}}
\end{center}
\end{table}

Before applying model testing or model checking, we first convert the textual requirements in the benchmark into their equivalent  STL formulas. Model checking approaches typically receive a temporal logic property and a model  as input.  For model testing, however,  we need to transform the logical properties into quantitative fitness functions (see Figure~\ref{fig:mc-mt}). To do so, we use a translation of STL into a robustness metric~\cite{fainekos2009robustness} which is summarized in Table~\ref{tbl:fitness}. The translation function $R$ is defined over a set $S = \{s_1, \ldots, s_n\}$ of signals at time $t$. We assume that the signals in $S$ are defined over the same time domain, i.e., for every $s_i \in S$, we have $s_i:[a,b] \rightarrow \mathcal{R}_i$ where $[a,b]$ is the common domain between signals in $S$. The choice of the $max$ and $min$ operators for defining the semantics of $\exists$ and $\forall$  is standard~\cite{larsen1988modal}: the minimum has the same  behavior as  $\wedge$  and evaluates  whether a predicate holds over the entire time  interval. Dually,  the max operator captures $\vee$.

For every STL property $\varphi$ and every set $S = \{s_1, \ldots, s_n\}$ of signals at time $t$, we have $R_{(S, t)}(\varphi) \geq 0$ if and only if $\varphi$ holds over the set $S$ at time $t$~\cite{fainekos2009robustness,Menghi:19}.  That is, we can infer boolean satisfaction of STL formulas based on their fitness values computed by $R$. In  Table~\ref{tbl:fitness},  $\epsilon$ is an infinitesimal positive value that is used to ensure the above relation between boolean satisfiability and fitness values of real-valued constraints (i.e., $\mu\; \textbf{rel-op}\; 0$)  and literals (i.e., $\top$, $\bot$, and $s^{\mathbb{B}}$) in the STL grammar.

We translate the requirements in the benchmark Simulink models into STL. Some examples of STL formulas corresponding to the requirements in our benchmark are shown in Table~\ref{tab:requirements}. For example, the formula $F_{[0,T]} G_{[0,T]} \left(  | \mathit{out} - \mathit{desired}| \le 0.01 \right)$ indicates that there is a time $t \in [0,T]$ such that for any time $t'$ such that $t'\geq t$, the constraint $ | \mathit{out}(t') - \mathit{desired}(t')| \le 0.01$ holds. As an example, the $R$ translation of this formula is given below: 

$\mbox{max}\big\{ \mbox{min} \big\{ 0.01 - | \mathit{out}(t') - \mathit{desired}(t')| \big\}_{t' \in [t, t+T]}  \big\}_{t \in [0,T]} $

To provide more detailed information about  the requirements in our benchmark, we present the  recurring \emph{temporal patterns} in the STL formulation of our benchmark requirements.  Table~\ref{tbl:pattern-temp} shows the temporal patterns  we identified in our study. The invariance pattern, which simply states that a property should hold all the time, is the most recurring temporal pattern in our Simulink model benchmark. The other patterns in  Table~\ref{tbl:pattern-temp} capture common controller requirements, i.e., stability or steady-state, responsiveness, smoothness, and fairness. Note that in  Table~\ref{tbl:pattern-temp}, the time interval for $G$ operators are expected to be the same as the time domain of the signals to which the operators are applied. In Table \ref{table:patterncm}, we show the list of the temporal patterns  appeared  in formalisations of the requirements of each model in our Simulink benchmark.  As this table illustrates, the invariance pattern (\textbf{T1}) is used for some requirements of every model. The other temporal patterns (i.e., \textbf{T2}, \textbf{T3}, \textbf{T4}, and \textbf{T5}) only appear in requirements formalisations of models that include some continuous controllers (i.e. Autopilot and Regulator). 

\begin{table}[t]
\caption{Temporal patterns in STL translations of our benchmark requirements.}
\label{tbl:pattern-temp}
\begin{center}
\scalebox{.7}{\begin{tabular}{|p{0.3\linewidth}|p{0.18\linewidth}|p{.8\linewidth}|}
\hline
\textbf{Name-ID} & \textbf{STL formulation} & \textbf{Explanation} \\
\hline
\textbf{Invariance - T1} & $G \varphi$ & The system should always exhibit the behaviour $\varphi$. \\
\hline
\textbf{Steady State - T2} & $F_{[0,d]} G \varphi$ & The system within the time duration $[0,d]$ exhibits the behavior $\varphi$ and continues  exhibiting this behavior until the end.   \\
\hline
\textbf{Smoothness - T3} & $G (\psi \Rightarrow G \varphi)$ &  Whenever the system exhibits $\psi$, it has to exhibit $\varphi$ until the end.  \\
\hline
\textbf{Responsiveness - T4} & $F_{[0,d]} \varphi$ & The system shall exhibit $\varphi$ within the time duration of $[0,d]$.  \\
\hline
\textbf{Fairness - T5}  & $G F_{[0,d]} \varphi$ & At every time $t$, it should be possible for the system to exhibit the behaviour $\varphi$  within the next time duration $[t , t+d]$. \\
\hline
\end{tabular}}
\end{center}
\end{table}

\begin{table}[t]
\caption{Temporal patterns used in the requirements formalisations of each Simulink benchmark model.} 
\label{table:patterncm}
\begin{center}
\scalebox{.75}{\begin{tabular}{|p{0.2\linewidth}| p{0.1\linewidth} |p{0.2\linewidth}|p{0.2\linewidth}| p{0.1\linewidth} |p{0.2\linewidth}|}
 \hline             
 \bf Model   & \bf \# Req & \bf  Patterns &
  \bf Model  & \bf \# Req & \bf  Patterns
 \\ [0.5ex] 
 \hline                  
 	 Autopilot & 11 & T1, T2, T3, T4 & 
  	Two Tanks & 32 & T1  \\\hline
   	Neural Network & 3 & T1  & 
 	Tustin & 5 & T1  \\\hline
	FSM & 13 & T1   & 
	Nonlinear Guidance & 2  & T1   \\\hline
	Regulator &  10 & T1, T5   & 
	Euler &  8 & T1\\\hline
	SWIM & 2  &T1    &	Effector Blender & 3 & T1  \\\hline
	Triplex & 4 & T1 & & &  \\
\hline
\end{tabular}}
\end{center}
\end{table}


\section{Our Model Checking Technique}
\label{sec:mc}
SMT-based model checking has a long history of application in testing and verification of  CPS models. Briefly, to check if a model $M$ meets its requirement $r$, the requirement is first translated into a logical property $\varphi$. An SMT-solver is then used to prove satisfiability of $M \wedge \neg \varphi$. If  $M \wedge \neg \varphi$ turns out to be SAT, then $M$ does not satisfy $\varphi$. If  $M \wedge \neg \varphi$ is UNSAT, it implies that $M$ satisfies $\varphi$. In general, SMT-based model checkers are  focused on checking safety properties (i.e., properties expressed using the $G$ temporal operator). The liveness properties (i.e., properties that use the $F$ temporal operator) can be expanded assuming that they are specified over a finite time interval. For example,  $F_{[0,d]} \varphi$  can be rewritten as $\bigvee_{t\in [0,d]} \varphi(t)$ assuming that $[0,d]$ is discrete time interval.

In our study,  for the reasons discussed in Section~\ref{sec:intro},  we use QVTrace as a representative SMT-based model checking tool for Simulink.  In addition to the standard SMT-based model checking described above, QVTrace uses the $k$-induction technique~\cite{DonaldsonHKR11} to enhance the set of formulas it can verify. QVTrace uses a logical predicate language referred to as QCT to  capture requirements.  QCT supports all the numerical and boolean operators described in STL grammar, but similar to most existing SMT-based model checkers, among the temporal operators, it only supports the  temporal operator $G$, i.e., globally.  Hence, among the temporal patterns in Table~\ref{tbl:pattern-temp}, QCT can specify \textbf{T1} and \textbf{T3} directly. Properties involving  \textbf{T2}, \textbf{T4} and \textbf{T5} can be expressed in QCT after we expand them as discussed earlier. Specifically, as we will discuss in Section~\ref{sec:eval},  the requirements that used temporal patterns   \textbf{T2}, \textbf{T4}  belong to  Autopilot  that could not be verified using QVTrace due to its complex features, and the requirement of the Regulator model that used the  \textbf{T5} pattern was expressed in QCT  as a large disjunctive formula (i.e., $G \bigvee_{t\in [0,d]} \varphi(t)$).  We note that, in general, while being a subset of STL, QCT is  sufficiently expressive for most  problems we have seen in practice. In addition, QCT is carefully designed to be easy to read and understand by a typical engineer who may not have background in temporal logic. Finally, there is an  efficient and straightforward translation from QCT into the input  languages of SMT-solvers and theorem provers.


When the SMT-based formulation of  $M \wedge \neg \varphi$ becomes so large that it cannot be handled by the underlying SMT-solvers,  QVTrace relies on bounded model checking (BMC)~\cite{ClarkeBRZ01}  mainly to identify inputs that falsify the model under test. BMC  checks the behavior of the underlying model for a fixed number of steps $k$ to see if a counter-example with length less than $k$ can be found for the property of interest. As a result, BMC  can only falsify properties up to a given depth $k$ and is not able to prove the correctness of the model with respect to a property.

\section{Our Model Testing Technique}
\label{sec:mt}
Recently, there has been a surge of interest in  using falsification methods for testing Simulink models~\cite{BriandNSB25,matinnejad2018test,Abbas:13}. These methods, which we refer to as model testing, are black-box and rely on simulation outputs (i.e., model executions) to 
generate test inputs that falsify the given requirements. Figure~\ref{fig:mc-mt}(a)  shows an overview of our model testing framework.   In our work, we use evolutionary search algorithms to generate test inputs falsifying a given requirement. Search algorithms sample the input space, selecting the fittest test inputs, i.e., test inputs that are (likely) violating or are as close as possible to violating  the requirement under analysis. Then they evolve the  fittest test inputs using genetic or evolutionary  operators to generate new test inputs and reiterate through the search loop~\cite{Luke2013}. The test inputs are expected to eventually move towards the fittest regions in the input space (i.e., the regions containing fault-revealing test inputs). This approach takes as input: (1)~the model under test, (2)~a fitness function guiding the search towards requirements violations, and (3)~the value ranges of the model input variables.  We discuss the fitness functions and the input search space  below. We then present a well-known evolutionary search algorithm used in our work.

Fitness functions  are computed based on the  model outputs obtained by running the model under test for sampled test inputs.  We use the robustness metric~\cite{fainekos2009robustness} as fitness functions in our work and use the translation in Table~\ref{tbl:fitness} to generate them from STL requirements formalizations. The robustness function $R(\varphi)$ is a value in $[-\infty, +\infty]$ such that $R(\varphi) \geq 0$ indicates that $\varphi$ holds over model outputs (and hence the test satisfies $\varphi$), and $R(\varphi) < 0$ shows that $\varphi$ is violated (and hence the test reveals a violation).  The robustness metric matches our notion of fitness as its value, when positive, shows how far a test input is from violating a requirement and when it is negative, its value shows how severe the failure revealed by a test is.


Since  model testing  works by sampling test inputs from the input search space of  the  model under test,  it requires the value ranges of the model input variables to be provided.  For each Simulink model in our benchmark, there is a document describing the model function and its requirements as well as its input and output variables. We extracted the value ranges of model input variables from these documents.

In this paper, we use a simple evolutionary search algorithm, known as \emph{hill climbing (HC)},  to generate test inputs (Algorithm~\ref{algo:hc}). This algorithm has been previously applied to testing Simulink models~\cite{MatinnejadNBBP15}.  
 The algorithm receives the search input space characterization $I$ and uses the fitness function $f$. It starts with randomly selected test input in the search space ($CS$ selected in line~2). It then iteratively modifies the test input (line~4), compares its fitness value with the fitness value of the best found test input (line~5), and replaces the best test input when a better candidate is found (line~6). The search continues until an optimal test input (i.e., yielding a negative fitness value) is found or we run out of the search time budget. The test inputs in our work are vectors of boolean, enum or real variables. Hence, we implement the $\mathit{Tweak}$ operator used in the HC algorithm  by applying a \emph{Gaussian  Convolution} operator~\cite{Luke2013} to the real variables and a \emph{Bit-Flip} operator~\cite{Luke2013} to the boolean and enum variables. The \emph{Bit-Flip} operator randomly toggles a boolean or an enum value to take another value from its range.  A \emph{Gaussian  Convolution} operator selects a value $d$ from a zero-centered Gaussian distribution ($\mu =0$, $\sigma^2$) and shift the variable to be mutated by the value of $d$. The  value of $\sigma^2$ is in the order of $0.005$ when we want to have an exploitative search operator (i.e., the one focused on locally search a small area of the search space) and is selected to be higher (e.g., more than $0.1$) when we are interested in more explorative search.

\begin{algorithm}
\caption{Hill Climbing Algorithm.}
\label{algo:hc}
\begin{algorithmic}[1]

\State  $I$ : Input Space
\State  $CS$ $\gets $initial candidate solution in $I$
\State \textbf{repeat }
	\State  \ \ \ \ $ NS$ $\gets$ Tweak(Copy(CS)) 
	 \State \ \ \ \ \textbf{if} $f(NS) < f(CS)$ \textbf{then} 
    	\State  \ \ \ \ \ \ \ \ \ $CS$  $\gets$ $NS $

\State \textbf{until }$CS$ is the ideal solution or we have run out of time
 \State \textbf{return }$CS$ 
 
\end{algorithmic}
\end{algorithm}

\section{Empirical Evaluation}
\label{sec:eval}
In this section, we report the results of applying the QVTrace tool (see Section~\ref{sec:mc}) and our model testing technique (see Section~\ref{sec:mt}) to our Simulink benchmark models described in Section~\ref{sec:cps-mod-req}. Specifically, we seek to answer the following research question:  \emph{How does model testing compare with (SMT-based) model checking in finding requirements violations in Simulink models?}

In the following, we explain the experimental setup we used for the evaluation.
Then, we answer our research question based on the results.




\subsection{Experiment Setup}
\label{subsec:exp}
As a prerequisite to apply both model testing and model checking to the benchmark Simulink models, we translated the textual requirements into STL (see Section~\ref{subsec:cpsreq}). We performed this translation in collaboration with our industry partner (the fifth and sixth authors of this paper). We had in total 92 requirements in our Simulink benchmark that we translated into STL. 
After that we used the translation in Table~\ref{tbl:fitness} to convert STL formulas into fitness functions to be used in our model testing approach. As discussed in Section~\ref{sec:mc}, we further translated STL properties into QCT, the property language of QVTrace.  We have made the benchmark Simulink models, their textual requirements and our STL translations of the requirements available online~\cite{AdditionalMaterial}.

After converting textual requirements into fitness functions and formal properties, we applied model testing and model checking to the models to identify requirements failures. In the model testing technique, we used the HC algorithm discussed in Section~\ref{sec:mt}. As discussed there, we used a  \emph{Gaussian Convolution} operator for the Tweak operation. In order for the HC search not to get stuck in   local optima, we opt for a relatively large value of  $\sigma^2$ for the Gaussian distribution from which the tweak values are chosen. Note that, in general, it is difficult to select a fixed value for $\sigma^2$ to tweak input variables of different models since these variables have different value ranges.  Hence, for each real-valued input variable $v$, we set $\sigma^2$ to be $0.1$ times the range width of $v$. We arrived at the value $0.1$ through experimentation.  If the tweaked values are out of variable ranges, we cap them at the max or min of the ranges when they are closer to the max or min, respectively. We set the main loop of HC  (see Algorithm~\ref{algo:hc}) to iterate for 150 times.  We chose this number because, in our preliminary experiments, the HC search has always reached a plateau after 150 iterations in our experiments. Finally,  in order to account for randomness in HC, for each model and for each requirement, we executed HC for 30 times. 

To apply QVTrace, we first investigate whether it is applicable to the given model. If so, then QVTrace attempts, in parallel, to exhaustively verify the property of interest or to identify input values falsifying the property. The former is typically performed using k-induction and the latter is done using bounded model checking (BMC). QVTrace generates four kinds of outputs: (1)~Green, when the property is exhaustively verified, (2)~Red, when  input values violating the property are found, (3)~Blue, when the property is verified upto a bound $k$, and (4)~Orange, when QVTrace fails to produce any conclusive results due to scalability or other issues.  In this paper, we present the results obtained based on the Z3 SMT solver~\cite{DeMoura:2008:ZES:1792734.1792766} since it had better performance than other solvers.

\begin{table*}[ht]
\caption{Comparing the abilities of model testing and model checking in finding requirements violations for Simulink models.} 
\label{table:sbtsmt}
\begin{center}
\scalebox{.67}{\begin{tabular}{|p{0.15\linewidth}| p{0.06\linewidth} |p{0.1\linewidth}| p{0.25\linewidth}| p{0.15\linewidth}| p{0.15\linewidth}|p{0.1\linewidth}| p{0.35\linewidth} |}
 \hline    
 \multirow{2}{*}{}{} & 
    \multirow{2}{*}{} & 
    \multicolumn{2}{c|}{\bfseries Model Testing (MT)} &
    \multicolumn{4}{c|}{\bfseries Model Checking (MC)}\\ \cline{3-8} 
    \bf {Model}&\bf{\# Reqs.}& \bf{\# Violations }&\bf \# Runs Revealing Violations &    {\bf \# Translated Reqs}&\bf{\# Proven Reqs }&\bf{\# Violations } &\bf{\#Proven Reqs using BMC up to the Bound $k$}  \\
 \hline                  
	Autopilot & 11 & 5 & $3(30/30), 1(4/30)$, $1(3/30)$&  0/11&  - & - &-\\\hline
  	Two Tanks & 32 & 11 & $10(30/30), 1(29/30)$ & 32/32 &  19 & 3 (11)* &10 ($k_1, \ldots k_8 \approx 90, k_9=110, k_{10}=260$) \\\hline 
   	Neural Network & 2 & 0&- & 2/2 &0& 0&2 ($k = 0$)\\\hline 
 	Tustin & 5 & 3 &$ 1(30/30), 1(29/30)$, $1(19/30)$& 5/5& 2&2  &1 ($k = 0$) \\\hline 
	FSM & 13 & 6 & $1(4/30), 1(6/30)$ $1(12/30), 1(9/30)$, $2(1/30)$&13/13&7& 6&0\\\hline 
	Nonlinear Guidance & 2  & 2 & $2(24/30)$  &2/2 &0&2 & 0   \\\hline 
	Regulator &  10 & 10 &$10(30/30)$& 10/10 & 0   & 9 &1 ($k = 110$)  \\\hline  
	Euler &  8 & 0  & - & 8/8 &  8&0  & 0\\\hline 
	SWIM & 2  & 0  & - &2/2 & 2&0 &0   \\\hline 
	Effector Blender & 3 & 2 & $ 1(30/30), 1(1/30)$ &3/3 &0 &0 & 3 ($k = 0$) \\\hline 
	Triplex & 4 & 1& $2(30/30)$ & 4/4 & 3 & 1 & 0 \\\hline \hline 
	\bf Total: & \bf 92 & \bf40 &\bf - & \bf81 & \bf41&\bf23&\bf17 \\

\hline
\end{tabular}}
\flushleft{\hspace*{.7cm}\small * QVTrace is able to find three violations when it is applied to the original Two Tanks model. If we modify the Two Tanks model to move the tanks' sensors closer together and to make the tanks smaller,  QVTrace is able to find the eleven violations  found by MT.   This is because violations are revealed much earlier in the simulation outputs of the modified Two Tanks model than in the outputs of the original model.}
\end{center}
\end{table*}

\subsection{Results}
Table~\ref{table:sbtsmt} reports the results of applying model testing and model checking to our Simulink model benchmark. Specifically, for model testing (MT), we report the number of requirements violations that we were able to find for each model. Recall that we executed HC 30 times for each requirement. Therefore, in Table~\ref{table:sbtsmt}, we report for each model and each violated requirement the number of fault revealing runs of MT.  For example, out of 11 requirements in Autopilot, MT is able to identify five requirements violations. Three of these violations  were revealed by 30 runs, one of them by four runs and the last by three runs. Since MT is black-box and analyzes  simulation outputs, it is applicable to any Simulink model that can be executed. That is, it is applicable to all the benchmark models and requirements.

For model checking (MC), for each model, we report whether or not the model or all the requirements of a model could be analyzed by QVTrace (i.e.,  if the models and requirements could be translated into an internal model to be passed into SMT solvers).   For the  models and requirements that could be analyzed by QVTrace, we report in Table~\ref{table:sbtsmt}:  (1)~the number of requirements that can be checked exhaustively and proven to be correct, (2)~the number of identified  requirements violations, and (3)~the number of requirements that were checked by bounded model checking (BMC) up to a specific bound $k$ for which no violation was found.

For example, as shown in Table~\ref{table:sbtsmt}, QVTrace was not able to translate the Autopilot model. This is indicated in the table by showing that 0 out of the 11 requirements of Autopilot could be translated internally by QVTrace.  However, QVTrace is able to handle Two Tanks and its 32 requirements. Among  these, QVTrace proves 19 requirements to be correct, finds three requirements violations and is able to check ten requirements using BMC up to the following  bounds, respectively: $k_1 \ldots k_8 \approx 90$,  $k_9 = 110$,  and $k_{10} = 260$. Specifically, for these ten requirements of Two Tanks, BMC is able to check the correctness of each requirement $r_i$ up to the depth $k_i$ (where $1 \leq i \leq 10$), but the underlying SMT-solver fails to produce results  for any depth $k > k_i$ due to scalability issues.  We note that, for Two Tanks, QVTrace is able to find all the 11 violations found by MT if the Two Tanks model is modified such that the tanks are smaller and the tanks' sensors are closer together.   This is because violations are revealed much earlier in the simulation outputs of the modified Two Tanks model than in the outputs of the original model. Finally, for some of the requirements of some models (i.e., Neural Network, Tustin and Effective Blender), BMC was not able to prove the requirements of interest for any bound $k$.  In Table~\ref{table:sbtsmt},  we use $k = 0$ in the BMC column to  indicate the cases where  the SMT-solver failed to produce any results for $k = 1$ when the BMC mode of QVTrace is used.

 
Table~\ref{table:timemcmt} compares the time performance of running MT and MC. On average, across all the models, each run of MT took 5.8min. The maximum average  execution time of an MT run (i.e., 150 iterations of the HC algorithm) was 18.5min (for Autopilot), and the minimum average execution time of MT was 3min (for Nonlinear Guidance). We note that the time required for running MT depends on the number of search iterations (which in our work is set to 150) as well as the time required to run a model simulation. The latter depends on the complexity of the model and  the length of the simulation time duration. Every Simulink model in the benchmark already has a default simulation time duration  that we used in our experiments.  

Proving each of the 41 requirements in the benchmark, which could be exhaustively checked by MC, took only 0.63sec on average.  The Two Tanks requirements required the longest average time to be proven (1.89sec), and the Euler requirements required the lowest average time to be proven (0.06sec). On average, it took MC 2.19sec to find 29 requirement violations in the benchmark. 
For the 17 requirements where BMC had to be tried, we have listed  the time it took for the BMC mode of QVTrace to report an ``inconclusive'' output when we try a bound $k$ larger than the maximum bound values that BMC could handle and are shown in  Table~\ref{table:sbtsmt}.  We note that as shown in Table~\ref{table:timemcmt}, there are variations in the time required by QVTrace to report ``inconclusive''. In particular, in some cases, it takes several minutes or even hours to report the ``inconclusive'' message and in some cases, the message is reported after a few seconds. This has to do with the internal choices made in QVTrace, but in either case, the ``inconclusive'' message indicates that  the underlying SMT-solver (i.e., Z3) is not able to report results either because the input to the SMT-solver is too large or because the solver cannot handle some features in its input.

\begin{table*}[ht]
\caption{Comparing the time performance of model testing and model checking.} 
\label{table:timemcmt}
\begin{center}
\scalebox{.7}{\begin{tabular}{|p{0.15\linewidth}| p{0.1\linewidth}| p{0.1\linewidth}| p{0.1\linewidth}| p{0.8\linewidth}|}
 \hline    
    \bf {Model}& {\bf avg. Time  per MT run}& {\bf avg. Time  to prove reqs (MC)}&\bf{avg. Time  to violate reqs (MC) } &\bf{BMC Time when QVTrace reports ``inconclusive'' for bound values $k$ larger than the ones reporter in Table~\ref{table:sbtsmt}}\\
 \hline                  
	Autopilot &  18.5min & - &- &-  \\\hline
  	Two Tanks & 5.1min  &1.89s &1.09s& For the ten requirements of Two Tanks that have to be checked by BMC, QVTrace reports  ``inconclusive'' after approximately 5min. \\\hline
   	Neural Network  &5.9min & -&-& QVTrace reports ``inconclusive'' for the two requirements of Neural Network after waiting for $1958.9s$ (32.6min) and $ 847.1s$ (14.1min), repectively. \\\hline
 	Tustin & 4.6min  & 0.19s&0.76s& QVTrace reports ``inconclusive'' for one requirement
	 of Tustin after waiting for $1121s$ (18.7min) .\\\hline
	FSM  &3.6min & 0.59s&0.18s& -\\\hline
	Nonlinear Guidance  & 3min&-& 0.12s& - \\\hline
	Regulator  & 3.6min& -&10.1s& QVTrace reports ``inconclusive'' for one requirement of Regulartor  after waiting for $1303.3$s (21.7min). \\\hline  
	Euler  &4.5min & 0.06s&-&-\\\hline
	SWIM   & 5.2min& 0.18s&-&- \\\hline
	Effector Blender  & 4.4min& -&-& QVTrace reports ``inconclusive'' for two requirements of Effector Blender  after waiting for
	$9475.4$s (2.6h) and $4371.9$s (1.2h), respectively.  For the third requirement of Effector Blender, QVTrace reports ``inconclusive'' after only $37.8$s.  \\\hline
	Triplex &5.6min & 0.88s &0.88s& -  \\\hline
	\hline
	{\bf Average}  & 5.8min& 0.63s&2.19s&-\\\hline

\end{tabular}}
\end{center}
\end{table*}

We note that all the requirements violations were communicated to Company A who developed the  benchmark and were confirmed as valid violation cases. The results show that all the violations discovered by MC were also discovered by MT, but there were violations that MT could discover that could not be identified by MC. Specifically, there were 17 violations that MT could find but not MC. Among these, five belonged to the Autopilot model that could not be handled by MC. The other 12 were among the 17 requirements that had to be checked by BMC, but BMC could not check the requirements beyond some bound $k$ while the failures could be revealed by MT  at a time step beyond  $k$. Finally, we note that  MC was able to exhaustively prove  41 requirements,  whereas MT, being a testing framework, is focused on fault-finding only. In Section~\ref{sec:ll}, we discuss the complementary nature of MT and MC  and will draw a few lessons learned based on our results.

\emph{In summary,} out of the 92 requirements in our benchmark, MT was able to identify 40 requirement violations and MC  only found 23 of them, without detecting additional violations.  Among the 40 violations found by MT, 32 were found by more than half of the runs. This shows, as we have seen before, that one should run MT as many times as possible. Among the 92 requirements, MC was able to prove correctness for 41 of them. Finally,  MC and MT together were able to either prove or find violations for 81  of the 92 requirements. 

\subsection{Data Availability} 
\label{subsec:data}
All the data, code and tool access required to replicate our study are available online. Specifically, we have made available online~\cite{AdditionalMaterial} the following: (1)~our Simulink benchmark including models, textual requirements and STL formalisations of the requirements. (2)~The implementation of our model testing technique  including fitness functions obtained based on the STL requirements formalisations. (3)~QCT descriptions of the requirements as well as instructions on how to access and use QVTrace using a docker virtual machine. (4)~Detailed reports on the results shown in Tables~\ref{table:sbtsmt} and \ref{table:timemcmt}.

\section{Lessons Learned}
\label{sec:ll}
We  draw five lessons learned based on our experiment results and our experience of applying MT and MC to the Simulink benchmark. Our aim is to identify strengths and weaknesses of the two techniques when they are used to verify Simulink models, and provide recommendations as to how MC and MT can be combined together to increase effectiveness of Simulink verification. 

\textbf{Lesson1:} {\it MC may fail to analyse some CPS Simulink models.}  As confirmed by QRA, the most serious obstacle in adoption of model checking tools by the CPS industry is that such tools may not be readily applicable to some industrial  Simulink models. In particular, the inapplicability issue is likely to happen for models capturing continuous and dynamical systems (e.g., Autopilot). Before one can apply a model checking tool, such models have to be decomposed into smaller pieces, their complex features have to be simplified and the black-box components may have to be replaced by concrete implementations. We note that Autopilot could be analyzed by QVTrace after  removing the wind subsystem and discretising some computations (e.g., by replacing $\int$ with sum or $df/dt$ with $\Delta f/ \Delta t$). However, such simplifications and modifications may not be always feasible because: (1)~The simplifications may modify the model behavior and may  compromise analysis of some system requirements. This undermines the precision of analysis performed by MC, and further, some system requirements that are related to the simplified or removed subsystems can no longer be checked by MC. (2)~Such changes are expensive and require additional effort that may not be justified in some development environments.


\textbf{Lesson2:} {\it Bounded model checking  may fail to reveal violations that can be, otherwise, easily identified by MT.} In our study, bounded model checking (BMC) has been successfully used for analysis of 17 requirements to which  model checking could not be applied exhaustively.   MT, however,  was able to reveal violations for 12 of these 17 requirements. All these violations were obviously  revealed at time steps greater than the selected bounds $k$ in BMC. For example, for Two Tanks, MT was able to violate  eight requirements that were proven to be correct by BMC up to a bound less than 270. But these violations could be revealed at  around  500 and 1000 time steps of Two Tanks outputs.

\textbf{Lesson3:} {\it MC executes considerably faster than MT when it can prove or violate requirements. However, MC may quickly fail to scale when models grow in size and complexity.} MC executes considerably faster than MT  when it can conclusively prove or violate a requirement and does not warrant the use of BMC. While it took MC less than a couple of seconds, on average, to prove properties or to find violations for the benchmark, the quickest run of MT took about 3min.  While for small models, MC is quicker  than  MT, this trend unlikely holds for larger and more complex models. In particular, MC has the worst performance for Autopilot, Neural Network and Effector Blender that have complex features such as  continuous dynamics, non-algebraic functions and machine learning components. Some of the limitations, however, are due to the underlying SMT-solvers. 


\textbf{Lesson4:} {\it MT approaches, though effective at finding violations, need to be made more efficient on large models.}
In this paper, we used a relatively simple model testing approach implemented based on a Hill-Climbing algorithm guided by existing fitness functions proposed in the literature. MT approaches can be  improved in several ways to increase their effectiveness and practical usability. In particular, MT is computationally expensive as it requires to run the underlying model a large number of times. Since different runs of MT are totally independent, an easy way to rectify this issue is to paralellize the MT runs, in particular,  given that  multicore computers are now a commodity. In addition, there are several strands of research that investigate different search heuristics or attempt to combine search algorithms with surrogate models to reduce their computational time (e.g.,~\cite{AbdessalemNBS16,MatinnejadNBB14}).

\textbf{Lesson5:} {\it More empirical research is required to better understand what search heuristics should be used for what types of models.} Engineers are provided with little information and guidelines as to how they should select search heuristics to obtain most effective results when they use MT.  Each run of MT samples and executes a large number of test inputs. The generated data, however, apart from guiding the search, is not used to draw any information about the degree of complexity of the search problem at hand or to provide any feedback to engineers as to whether they should keep running MT further  or whether they should modify the underlying heuristics of their search algorithms. We believe further research is needed in this direction to make MT more usable and more effective in practice.

{\bf Combining MC and MT.} Our experience shows that MT and MC are complementary. MC can effectively prove the correctness of requirements when it is able to handle the size and the complexity of the underlying models and properties, while MT is effective in finding requirements violations. Indeed, for our benchmark, MC and MT together are able to prove 41 requirements and find 40 violations, leaving only 11 requirements (i.e.,\%12) inconclusive.  Given that MC is quite fast in proving and violating requirements, we can start by applying MC first and then proceed with MT when models or requirements cannot be handled by MC or its underlying SMT-solvers.

\section{Conclusions}
\label{sec:con}
In this paper, we presented an industrial Simulink model benchmark and used this benchmark to evaluate and compare capabilities of model checking and model testing techniques for finding requirements violations in Simulink models. Our results show that  our model checking technique  is effective and efficient in proving correctness of requirements on Simulink models that represent CPS components. However, as Simulink models become larger and more complex, in particular, when they involve complex non-algebraic or machine-learning components  or exhibit continuous dynamic behaviour, it becomes more likely that model checking or bounded model checking
fail to handle them or identify faults in them.  On the other hand, while our model testing technique can scale  to large and complex CPS models and identify some of their faults,  it  is still computationally expensive and does not provide any quidelines on what search heuristics should be used for what types of models. In the end, we believe combing the two techniques is the best way ahead. We also believe more studies comparing the performance of these techniques in different contexts can help researchers better identify limitations and strengths of these two main-stream automated verification techniques.


\bibliographystyle{ACM-Reference-Format}
\bibliography{bibliography}

\end{document}